\documentclass[%
preprint,
showpacs,
amsmath,amssymb,
aps,
pra,
]{revtex4-1}

\usepackage[super]{nth}
\usepackage{textcomp}
\usepackage{graphicx}

\begin{document}

\preprint{}

\title{Reduction of the low-temperature bulk gap in samarium hexaboride under high magnetic fields}

\author{S. Wolgast}
\email{swolgast@umich.edu}
\author{Y. S. Eo}
\author{K. Sun}
\author{\c{C}. Kurdak}
\affiliation{Department of Physics, University of Michigan, Ann Arbor, Michigan 48109-1040, USA}

\author{F. F. Balakirev}
\author{M. Jaime}
\affiliation{National High Magnetic Field Laboratory, Los Alamos National Laboratory, MS-E536, Los Alamos, New Mexico 87545, USA}

\author{D.-J. Kim}
\author{Z. Fisk}
\affiliation{University of California at Irvine, Dept.~of Physics and Astronomy, Irvine, California 92697, USA}

\date{\today}

\begin{abstract}
SmB$_6$ exhibits a small (15-20 meV) bandgap at low temperatures due to hybridized $d$ and $f$ electrons, a tiny (3 meV) transport activation energy ($E_{A}$) above 4 K, and surface states accessible to transport below 2 K. We study its magnetoresistance in 60-T pulsed fields between 1.5 K and 4 K. The response of the nearly $T$-independent surface states (which show no Shubnikov--de Haas oscillations) is distinct from that of the activated bulk. $E_{A}$ shrinks by 50\% under fields up to 60 T. Data up to 93 T suggest that this trend continues beyond 100 T, in contrast with previous explanations. It rules out emerging theories to explain observed exotic magnetic quantum oscillations.
\end{abstract}


\maketitle

\section{Introduction}
Samarium hexaboride (SmB$_6$) has a long, rich history as a paradigm mixed valent insulator, and it has recently enjoyed renewed scientific interest due to its predicted \cite{Dzero2010, Takimoto, Lu} behavior as a three-dimensional (3D) time-reversal-invariant topological Kondo insulator (TKI) and experimental confirmations \cite{Wolgast, Kim13,Zhang,Syers} of robust surface states consistent with the TKI picture. The opening of a small bulk bandgap at low temperatures is responsible for a remarkably robust insulating bulk behavior, and it has been the subject of numerous studies \cite{Allen,Cooley95,Cooley99,Denlinger} over the years as researchers seek to understand its formation mechanism due to strong correlation interactions. Recent de Haas--van Alphen (dHvA) measurements \cite{Lu,Suchitra2015} suggest that the bulk gap may be much more exotic than previously realized. In particular, the interpretation of Tan \textit{et al.} \cite{Suchitra2015} is for sizeable Fermi surfaces paradoxically coincident with the charge gap (\textit{cf.} Ref.~\cite{Li} for an alternative interpretation). New theories are currently emerging to explain this paradox \cite{KnolleCooper2015,Wang2016,Erten2016,ThomsonSachdev2016,Montambaux2016}, some of which (\textit{e.g.}, an oscillating bandgap scenario) are ruled out by our data here.

By now, there is experimental consensus for two distinct bandgap energies, depending on the type of probe used. Photoemission spectroscopy, optical spectroscopy, and tunneling spectroscopy all reveal a bandgap that opens around 150 K and widens to between 14 and 25 meV at temperatures below 10 K. One angle-resolved photoemission spectroscopy (ARPES) study \cite{Denlinger} reports a detailed temperature evolution in which the conduction band shifts down across the Fermi energy as the temperature is raised above 50 K. Meanwhile, transport measurements \cite{Menth,Allen,Dressel1999,Flachbart,FlachbartSpect2001,Gabani2001,Wolgast} yield tiny activation energies between 2 and 4 meV at temperatures between 5 and 20 K. Possible explanations for the difference between these values include an indirect bandgap scenario or the presence of in-gap bulk states or surface states pinning the Fermi energy a few meV below the conduction band \cite{Flachbart}. Although ARPES has unambiguously revealed the presence of two-dimensional (2D) surface states within the gap, the topological character of these states is still debated due to ARPES's limited energy resolution within the small gap \cite{Denlinger,Neupane,Miyazaki,Jiang,XuSARPES,NXu,Rader}. In the exotic scenario where there are no in-gap bulk states, it is a possibility that such surface states can play a role in pinning the Fermi energy so far from the band center \cite{AlexaDepletionWidth}. Evidence for 3D bulk in-gap states is much less clear, especially since most of the relevant studies \cite{Dressel1999,FlachbartSpect2001,Gabani2001,Nozawa2002} associated the signatures of in-gap states with the metallic behavior below 2 K, which is now known to be caused by the 2D surface states in the gap. The question of Fermi-level pinning remains unresolved.

One regime that is inaccessible to ARPES studies is that of high magnetic fields. Understanding how the bandgap and any in-gap states evolve with magnetic field can give insight regarding the underlying physics of the gap formation and structures within the gap. The historical benchmark transport study of SmB$_6$ by Cooley \textit{et al.} \cite{Cooley99} observed strong negative magnetoresistance (MR) at 4 K up to 60 T pulsed field. In an explosive flux measurement in the same study up to 142 T, the resistance reaches 1.5\% of its zero-field value around 86 T, after which the sample shows strong quadratic positive MR. These researchers, unaware of the onset of surface-dominated conduction at this temperature, attributed the negative MR and its minimum to the closure of the bulk bandgap around 86 T. More recently, a systematic temperature-dependent MR study \cite{Flachbart2009} also reached a similar conclusion. Unfortunately, this study does not provide much detail or explanation on the MR below 5 K, which we now know to be dominated by surface conduction \cite{Wolgast,Kim13}. This leaves open the question whether the MR signatures in this temperature range really correspond to a bulk behavior (\textit{e.g.}, a reduction of the bandgap) or a magnetically-induced enhancement of the surface conduction.

Motivated by the desire to separate the bulk MR behavior from the surface MR behavior we observed previously \cite{WolgastMR}, as well as to detect 2D transport signatures of the surface states (\textit{e.g.}, Shubnikov--de Haas (SdH) oscillations), we have conducted temperature-dependent transport measurements up to 60 T from 1.4 K to 4.0 K, along with measurements up to 93 T at 1.4 K. The temperature dependence provides a quantitative measure of the activation energy's ($E_{A}$) dependence on the magnetic field, while clearly distinguishing it from the behavior of the surface states. We conclude that the strong negative MR observed by Cooley \textit{et al.} is indeed due to the reduction of the bulk $E_{A}$ between the band edge and the Fermi energy, and is distinct from the smaller negative MR displayed by the surface states at even lower temperatures ($<2$ K), as reported by older \cite{Sugiyama1988} and more recent studies \cite{WolgastMR,Thomas,Nakajima}. However, we estimate that $E_{A}$ does not shrink to zero until around 120 T, much higher than estimated by Cooley \textit{et. al} \cite{Cooley99}. Meanwhile, the surface MR shows no indications of SdH oscillations up to 93 T, presumably due to the low mobilities of the surface states \cite{WolgastMR}. This result is mysterious in light of magnetization measurements \cite{Lu,Suchitra2015}, which clearly show quantum oscillations as low as 3 - 4 T.

\section{Experimental Methods}

We performed resistance measurements on Corbino disks fabricated on single (001) and (011) crystallographic surfaces of single-crystal SmB$_6$ grown via the aluminum flux method. The surfaces were prepared by lapping and polishing with Al$_2$O$_3$ slurry down to 0.3 $\mu$m grit size, and Corbino disks with inner diameter of 300 $\mu$m and outer diameter of 500 $\mu$m were fabricated using standard photolithography. 20/1500 \AA{} Ti/Au contacts were deposited by evaporation to form the metalized portion of the disks, and they were wirebonded using wedge bonding.

The Corbino disks' resistances were measured in one of the National High Magnetic Field Laboratory's  (NHMFL's) 65-T short pulse magnets using a standard resistance bridge and specialized AC (274.5 kHz) lock-in techniques developed for NHMFL's pulsed fields. The resistance depends on the sample geometry, which here may vary between 2D and 3D current paths; we thus report the raw resistance in the figures, which may be converted to sheet resistivity using $R_{\Box}=R\frac{2\pi}{\ln(b/a)}\approx12.30R$, where $a$ and $b$ are the inner and outer radii of the Corbino disk, respectively. The samples were immersed in liquid $^4$He at various pressures to obtain stable temperatures from approximately 1.5 to 4 K. At each temperature, a series of magnetic field pulses with incrementally larger maximum fields (and thus larger d$B$/d$t$) was used to obtain traces of resistance versus field, and to verify that d$B$/d$t$ effects (such as heating and inductive pickup) were not significant. Down-traces obtained from one (001) sample during the 60-T pulses are shown in Fig.~\ref{fig:Data}(a) (the up-traces include d$B$/d$t$ effects at the onset of the pulse). Additional data were taken up to 95 T at 1.4 K in the NHMFL's 100-T pulsed magnet using the same cryo-insert; this trace faithfully reproduces the 60-T trace at 1.54 K. The 95-T data, together with the 60-T trace at 3.96 K, are plotted as normalized MR with Cooley \textit{et al.}'s data in Fig.~\ref{fig:Data}(b).

\section{Results and Discussion}

\begin{figure}
\begin{center}
	\includegraphics{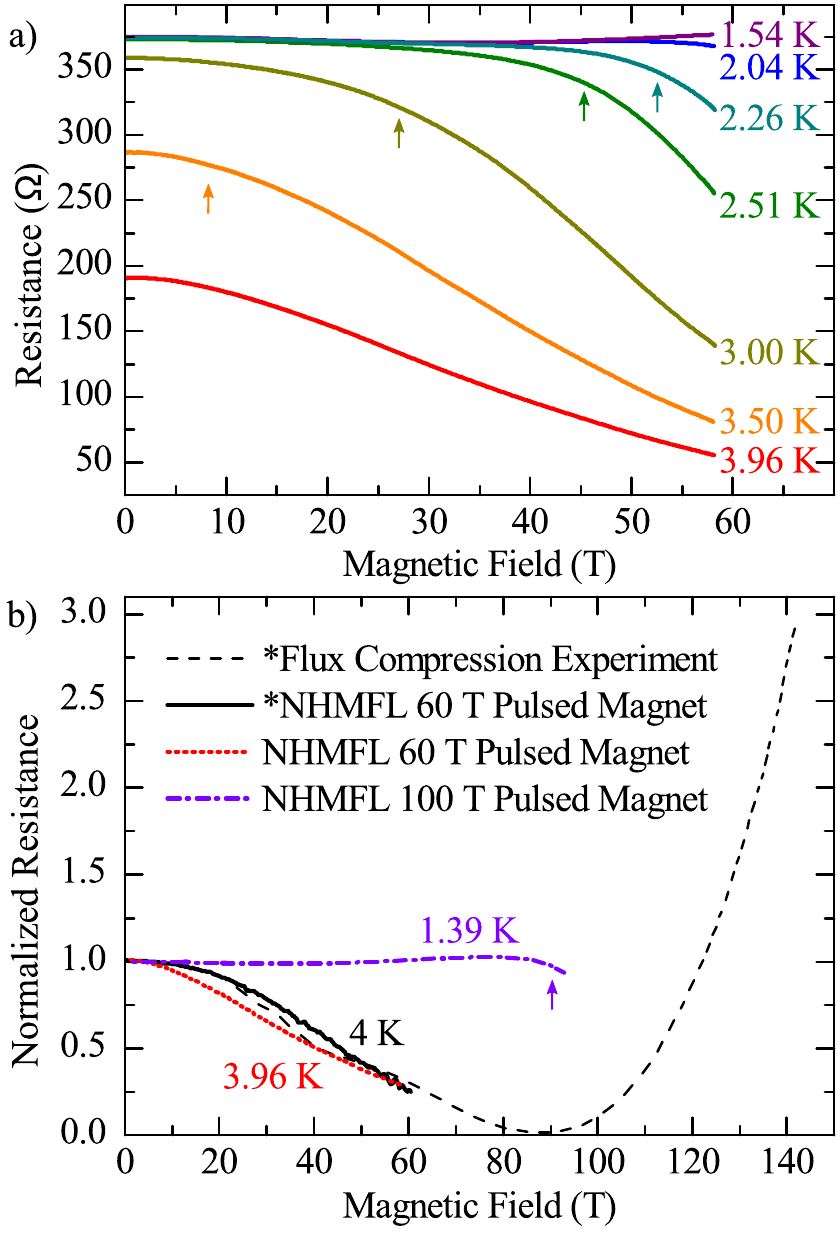}
	\caption{(Color online) (a) Traces of the Corbino disk two-terminal resistance as a function of magnetic field at various temperatures between 1.54 and 3.96 K. Arrows roughly indicate the positions of shoulders in the traces. (b) Traces of the Corbino disk two-terminal normalized resistance at 1.39 K and 3.96 K, plotted together with Cooley \textit{et al.}'s data. Cooley \textit{et al.}'s data is denoted by asterisks. The arrow roughly indicates the position of the shoulder in the 1.39 K trace.}
	\label{fig:Data}
\end{center}
\end{figure}

A few features of the MR data are immediately apparent. As the temperature is reduced, the zero-field value increases, then plateaus, as has been observed in virtually every DC resistance measurement of SmB$_6$, corresponding to a crossover from bulk-dominated to surface-dominated transport. The MR trace taken at 3.96 K is a reasonably faithful reproduction of the data taken by Cooley \textit{et al.} at 4 K using the NHMFL 65-T pulsed magnet \cite{Cooley99} (Fig.~\ref{fig:Data}(b)). At lower temperatures, the ``shoulder'' (indicated roughly by arrows in Fig.~\ref{fig:Data}(a)) associated with the crossover shifts to higher field values. At 1.54 K, no shoulder is discernable up to 60 T. In fact, at 1.39 K (Fig.~\ref{fig:Data}(b)), the shoulder does not begin to appear until about 90 T. The traces taken at 2.51 K and below (Fig.~\ref{fig:Data}(a)) all lie together at low fields, and branch away from the lowest temperature trace at sequentially higher fields as the temperature is reduced. This indicates that the surface state is largely temperature-independent at this scale, and that its behavior is mostly independent of the bulk transport behavior over these temperatures. This picture is not so clear from prior MR reports \cite{Sugiyama1988,Cooley99,Flachbart2009}, especially since those MR traces are plotted in a normalized fashion. Our data around 1.5 K differ slightly across multiple samples and from previous data taken at the same temperatures \cite{Sugiyama1988,WolgastMR}, perhaps due to the Corbino disk geometry and variations in surface quality. Still, in all the reports considered here, the surface state MR is very small (up to 15\%) compared to the 4-K bulk MR. In fact, we can now understand the behavior of the normalized MR magnitude plotted in Ref.~\cite{Flachbart2009} as a primarily geometric effect in which the MR is exaggerated at the crossover temperature by the change in the current path. We also note that none of the samples we measured up to 93 T at 1.4 K show any signs of SdH oscillations, though this is not surprising, considering the small surface mobilities reported for similar samples \cite{WolgastMR,Syers}.

\begin{figure}
\begin{center}
	\includegraphics{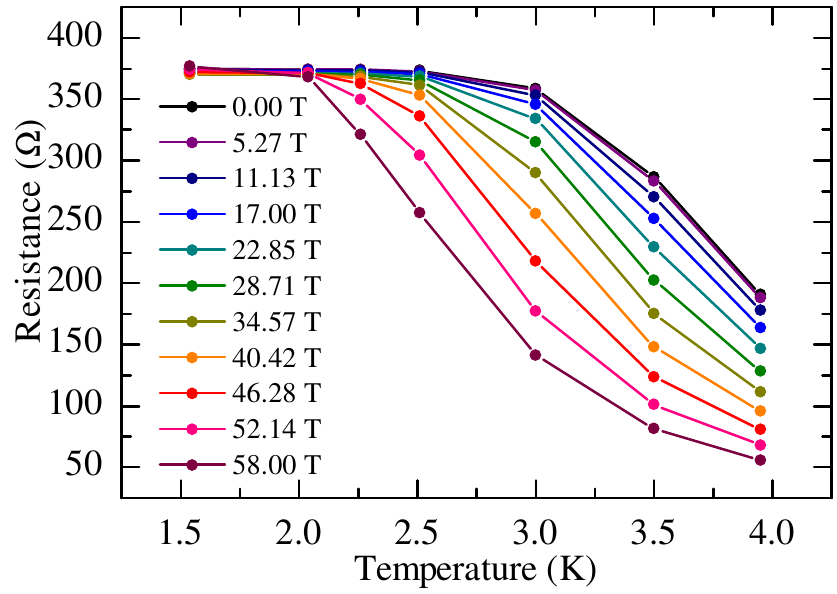}
	\caption{(Color online) Plots of the Corbino disk two-terminal resistance versus temperature for select values of magnetic field.}
	\label{fig:RvsT}
\end{center}
\end{figure}

The data in Fig.~\ref{fig:Data}(a) is replotted as resistance versus temperature in Fig.~\ref{fig:RvsT} by taking slices of the data at selected magnetic fields. This format visualizes the evolution of the resistance rise and plateau with the magnetic field strength. The crossover shoulder moves to lower temperatures as the magnetic field is increased.

\begin{figure}
\begin{center}
	\includegraphics{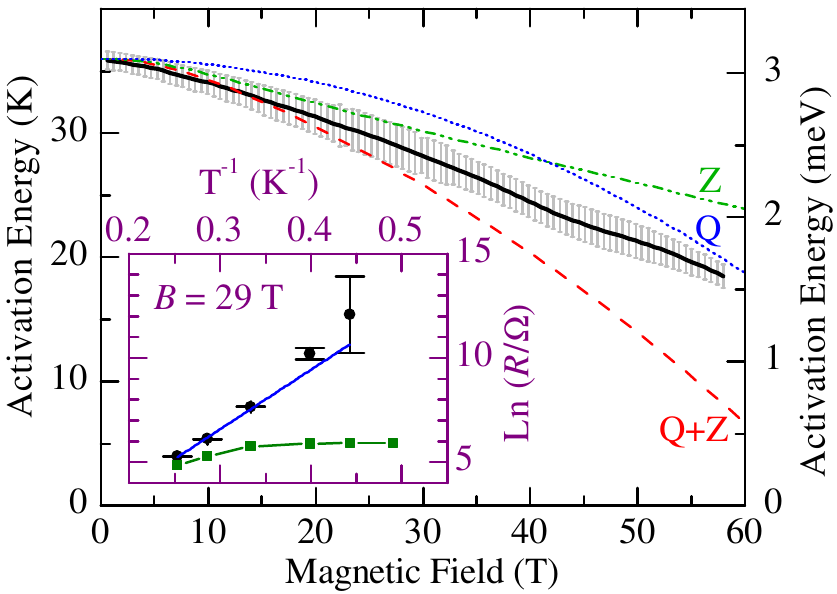}
	\caption{(Color online) Calculated transport $E_{A}$ (black line) as a function of magnetic field (error bars correspond to fit residuals). The data may be approximated by $E_{A}/\text{K}=35.77+0.01774 x-0.03587 x^2+0.00229 x^3-7.58845\times10^{-5} x^{4}+1.18699\times10^{-6} x^5-6.95936\times10^{-9} x^6$. Theoretical fits are also plotted for the Zeeman second-order shift, Zeeman splitting, and shift + splitting (blue dotted, green dash-dotted, and red dashed, respectively). Inset: Example weighted fit (blue line) at 29 T of the calculated bulk resistance data (black dots with error bars) on an Arrhenius plot. The raw resistance is also plotted (green squares) for reference.}
	\label{fig:ActivationEnergy}
\end{center}
\end{figure}

The temperature dependence can be used to estimate the size of the bulk transport gap at various magnetic fields. We model the total resistance as the parallel combination of the (temperature-independent) surface resistance and an activated bulk resistance:
$$R(B,T)=\left(\frac{1}{R_s(B)}+\frac{1}{R_0(B)}e^{\frac{\Delta(B)}{k_{\text{B}}T}}\right)^{-1},$$
where $R_s(B)$ is the surface resistance (taken from the trace at 1.54 K), and $R_0(B)$ and $\Delta(B)$ (the bandgap) are fitting parameters. We subtract off the contribution from the surface resistance and plot the remaining bulk contribution as an Arrhenius plot. Linear fits of this plot for each magnetic field value provide $E_{A}$ with uncertainty estimated by the fit residuals. The linear fits are weighted by the relative contribution of the bulk portion to the total resistance $\frac{\partial\ln(R_{b})}{\partial R}$, since the uncertainty in $R_{b}$ becomes large when $R_{b}\ll R_{s}$ ($R\approx R_{s}$). Figure~\ref{fig:ActivationEnergy} shows the best-fit $E_{A}$ versus magnetic field, with the uncertainty of $\Delta(B)$ indicated by the gray error bars. The inset depicts an example fit at 29 T, plotted with the raw resistance and the calculated bulk resistance with error bars. The transport gap closes with magnetic field to about 50\% of its zero-field value at 58 T. Although we do not have temperature-dependent data at higher fields, the onset of the crossover shoulder around 90 T at 1.39 K suggests that the closure continues to be approximately linear, reaching roughly 10 K at 93 T. Extrapolating the value of the gap closure is speculative at best, but such a linear trend suggests that the gap closes around 120 T. This data has some qualitative agreement with prior reports \cite{Chen2015} of field-induced gap closure. In any case, the finite resistance of the 93-T trace at 1.39 K alone is sufficient to reveal that the minimum at 86 T of Cooley \textit{et al.}'s 4-K flux compression data does not correspond to a fully-closed gap. Our measurements at this temperature are not at large enough fields to observe the positive MR in Cooley \textit{et al.}'s data above 86 T.

The gap seems to close in a nonlinear fashion, in qualitative agreement with Cooley \textit{et al.}'s data. Arguments for this qualitative behavior have been suggested elsewhere \cite{Cooley99,Flachbart2009}, but no descriptive models have been proposed. The identification by ARPES \cite{Denlinger} of the X-point of the conduction band as the band edge responsible for the small transport bandgap in this temperature range implies that the conduction band edge is driven toward the Fermi energy (or vice versa) under the influence of the magnetic field. We expect two mechanisms may be involved---Zeeman splitting of the Sm$^{3+}$ state and a second-order Zeeman shift of the Sm$^{2+}$ state. Alongside $E_{A}$ plotted in Fig.~\ref{fig:ActivationEnergy}, we plot the simulated behavior of the extracted $E_{A}$ for Zeeman splitting (Z) with a Land\'{e} $g$-factor of 0.6 \cite{WellsThesis,Wells1999,Wells2000}, for a second-order Zeeman shift (Q) of 1.0 Hz/G$^{2}$ \cite{Macfarlane1984}, and for the combination of these two effects (Q${}+{}$Z), which might be expected for the quasiparticle transition between Sm$^{2+}$ and Sm$^{3+}$ states (none of these contain any adjustable parameters besides the 0-field resistance). The data at low fields fit the combined case (Q${}+{}$Z) quite well. However, the data at high fields do not fit any of these models particularly well, especially the combined model which predicts the full closure of the gap around 74 T. These models assume, however, that the Fermi energy itself does not shift with magnetic field, which is a possibility suggested by earlier MR measurements of a shift in the surface states' carrier densities \cite{WolgastMR}. Also, there may be an additional second-order Zeeman shift associated with the Sm$^{3+}$ state, which unfortunately cannot be distinguished from the linear splitting at the low fields used so far to determine $g$. Thus, the quantitative results from these models are only valid for low fields. Meanwhile, the relation of the gap reduction to that observed with increasing pressure \cite{Cooley95} is unknown; whereas the gap suddenly collapses at pressures around 50 kbar, there is no collapse over the range of magnetic fields explored.

In summary, we have explored the MR of SmB$_6$ near the surface--bulk crossover temperature using large pulsed fields up to 93 T. We find that the surface states and the bulk give distinct contributions to the MR, and that the crossover follows a smooth curve in the $B$--$T$ plane. The surface states display a weak, sample-dependent MR, and we have not observed any SdH oscillations up to 93 T. Meanwhile, $E_{A}$ of the bulk carriers shrinks monotonically under large magnetic fields, but does not disappear until at least 93 T, in contrast with previous interpretations; extrapolation suggests the Fermi energy crosses into the conduction band around 120 T. An oscillating gap scenario used to explain the magnetic quantum oscillations is ruled out by the smooth, monotonic closure of the gap observed here.

\begin{acknowledgments}
We thank Piers Coleman and Jim Allen for their comments regarding bandgap closure mechanisms, and Yates Coulter for his assistance with the NHMFL 100-T pulsed magnet. The samples were prepared in part at the Lurie Nanofabrication Facility, a member of the National Nanotechnology Infrastructure Network, which is supported by the NSF. A portion of this work was performed at the National High Magnetic Field Laboratory, which is supported by National Science Foundation Cooperative Agreement \#DMR-1157490 and the State of Florida. Work at LANL was supported by the US DOE Basic Energy Science project ``Science at 100 Tesla.'' Funding was provided by NSF grants \#DMR-1006500, \#DMR-1441965, and \#DMR-0801253.
\end{acknowledgments}



\begin{thebibliography}{41}%
\makeatletter
\providecommand \@ifxundefined [1]{%
 \@ifx{#1\undefined}
}%
\providecommand \@ifnum [1]{%
 \ifnum #1\expandafter \@firstoftwo
 \else \expandafter \@secondoftwo
 \fi
}%
\providecommand \@ifx [1]{%
 \ifx #1\expandafter \@firstoftwo
 \else \expandafter \@secondoftwo
 \fi
}%
\providecommand \natexlab [1]{#1}%
\providecommand \enquote  [1]{``#1''}%
\providecommand \bibnamefont  [1]{#1}%
\providecommand \bibfnamefont [1]{#1}%
\providecommand \citenamefont [1]{#1}%
\providecommand \href@noop [0]{\@secondoftwo}%
\providecommand \href [0]{\begingroup \@sanitize@url \@href}%
\providecommand \@href[1]{\@@startlink{#1}\@@href}%
\providecommand \@@href[1]{\endgroup#1\@@endlink}%
\providecommand \@sanitize@url [0]{\catcode `\\12\catcode `\$12\catcode
  `\&12\catcode `\#12\catcode `\^12\catcode `\_12\catcode `\%12\relax}%
\providecommand \@@startlink[1]{}%
\providecommand \@@endlink[0]{}%
\providecommand \url  [0]{\begingroup\@sanitize@url \@url }%
\providecommand \@url [1]{\endgroup\@href {#1}{\urlprefix }}%
\providecommand \urlprefix  [0]{URL }%
\providecommand \Eprint [0]{\href }%
\providecommand \doibase [0]{http://dx.doi.org/}%
\providecommand \selectlanguage [0]{\@gobble}%
\providecommand \bibinfo  [0]{\@secondoftwo}%
\providecommand \bibfield  [0]{\@secondoftwo}%
\providecommand \translation [1]{[#1]}%
\providecommand \BibitemOpen [0]{}%
\providecommand \bibitemStop [0]{}%
\providecommand \bibitemNoStop [0]{.\EOS\space}%
\providecommand \EOS [0]{\spacefactor3000\relax}%
\providecommand \BibitemShut  [1]{\csname bibitem#1\endcsname}%
\let\auto@bib@innerbib\@empty
\bibitem [{\citenamefont {Dzero}\ \emph {et~al.}(2010)\citenamefont {Dzero},
  \citenamefont {Sun}, \citenamefont {Galitski},\ and\ \citenamefont
  {Coleman}}]{Dzero2010}%
  \BibitemOpen
  \bibfield  {author} {\bibinfo {author} {\bibfnamefont {M.}~\bibnamefont
  {Dzero}}, \bibinfo {author} {\bibfnamefont {K.}~\bibnamefont {Sun}}, \bibinfo
  {author} {\bibfnamefont {V.}~\bibnamefont {Galitski}}, \ and\ \bibinfo
  {author} {\bibfnamefont {P.}~\bibnamefont {Coleman}},\ }\href {\doibase
  10.1103/PhysRevLett.104.106408} {\bibfield  {journal} {\bibinfo  {journal}
  {Phys. Rev. Lett.}\ }\textbf {\bibinfo {volume} {104}},\ \bibinfo {pages}
  {106408} (\bibinfo {year} {2010})}\BibitemShut {NoStop}%
\bibitem [{\citenamefont {Takimoto}(2011)}]{Takimoto}%
  \BibitemOpen
  \bibfield  {author} {\bibinfo {author} {\bibfnamefont {T.}~\bibnamefont
  {Takimoto}},\ }\href {\doibase 10.1143/JPSJ.80.123710} {\bibfield  {journal}
  {\bibinfo  {journal} {J. Phys. Soc. Jap.}\ }\textbf {\bibinfo {volume}
  {80}},\ \bibinfo {pages} {123710} (\bibinfo {year} {2011})}\BibitemShut
  {NoStop}%
\bibitem [{\citenamefont {Lu}\ \emph {et~al.}(2013)\citenamefont {Lu},
  \citenamefont {Zhao}, \citenamefont {Weng}, \citenamefont {Fang},\ and\
  \citenamefont {Dai}}]{Lu}%
  \BibitemOpen
  \bibfield  {author} {\bibinfo {author} {\bibfnamefont {F.}~\bibnamefont
  {Lu}}, \bibinfo {author} {\bibfnamefont {J.~Z.}\ \bibnamefont {Zhao}},
  \bibinfo {author} {\bibfnamefont {H.}~\bibnamefont {Weng}}, \bibinfo {author}
  {\bibfnamefont {Z.}~\bibnamefont {Fang}}, \ and\ \bibinfo {author}
  {\bibfnamefont {X.}~\bibnamefont {Dai}},\ }\href {\doibase
  10.1103/PhysRevLett.110.096401} {\bibfield  {journal} {\bibinfo  {journal}
  {Phys. Rev. Lett.}\ }\textbf {\bibinfo {volume} {110}},\ \bibinfo {pages}
  {096401} (\bibinfo {year} {2013})}\BibitemShut {NoStop}%
\bibitem [{\citenamefont {Wolgast}\ \emph {et~al.}(2013)\citenamefont
  {Wolgast}, \citenamefont {Kurdak}, \citenamefont {Sun}, \citenamefont
  {Allen}, \citenamefont {Kim},\ and\ \citenamefont {Fisk}}]{Wolgast}%
  \BibitemOpen
  \bibfield  {author} {\bibinfo {author} {\bibfnamefont {S.}~\bibnamefont
  {Wolgast}}, \bibinfo {author} {\bibfnamefont {{\c{C}}.}~\bibnamefont
  {Kurdak}}, \bibinfo {author} {\bibfnamefont {K.}~\bibnamefont {Sun}},
  \bibinfo {author} {\bibfnamefont {J.~W.}\ \bibnamefont {Allen}}, \bibinfo
  {author} {\bibfnamefont {D.-J.}\ \bibnamefont {Kim}}, \ and\ \bibinfo
  {author} {\bibfnamefont {Z.}~\bibnamefont {Fisk}},\ }\href {\doibase
  10.1103/PhysRevB.88.180405} {\bibfield  {journal} {\bibinfo  {journal} {Phys.
  Rev. B (R)}\ }\textbf {\bibinfo {volume} {88}},\ \bibinfo {pages} {180405}
  (\bibinfo {year} {2013})}\BibitemShut {NoStop}%
\bibitem [{\citenamefont {Kim}\ \emph {et~al.}(2013)\citenamefont {Kim},
  \citenamefont {Thomas}, \citenamefont {Grant}, \citenamefont {Botimer},
  \citenamefont {Fisk},\ and\ \citenamefont {Xia}}]{Kim13}%
  \BibitemOpen
  \bibfield  {author} {\bibinfo {author} {\bibfnamefont {D.~J.}\ \bibnamefont
  {Kim}}, \bibinfo {author} {\bibfnamefont {S.}~\bibnamefont {Thomas}},
  \bibinfo {author} {\bibfnamefont {T.}~\bibnamefont {Grant}}, \bibinfo
  {author} {\bibfnamefont {J.}~\bibnamefont {Botimer}}, \bibinfo {author}
  {\bibfnamefont {Z.}~\bibnamefont {Fisk}}, \ and\ \bibinfo {author}
  {\bibfnamefont {J.}~\bibnamefont {Xia}},\ }\href {\doibase 10.1038/srep03150}
  {\bibfield  {journal} {\bibinfo  {journal} {Sci. Rep. -UK}\ }\textbf
  {\bibinfo {volume} {3}},\ \bibinfo {pages} {3150} (\bibinfo {year}
  {2013})}\BibitemShut {NoStop}%
\bibitem [{\citenamefont {Zhang}\ \emph {et~al.}(2013)\citenamefont {Zhang},
  \citenamefont {Butch}, \citenamefont {Syers}, \citenamefont {Ziemak},
  \citenamefont {Greene},\ and\ \citenamefont {Paglione}}]{Zhang}%
  \BibitemOpen
  \bibfield  {author} {\bibinfo {author} {\bibfnamefont {X.}~\bibnamefont
  {Zhang}}, \bibinfo {author} {\bibfnamefont {N.~P.}\ \bibnamefont {Butch}},
  \bibinfo {author} {\bibfnamefont {P.}~\bibnamefont {Syers}}, \bibinfo
  {author} {\bibfnamefont {S.}~\bibnamefont {Ziemak}}, \bibinfo {author}
  {\bibfnamefont {R.~L.}\ \bibnamefont {Greene}}, \ and\ \bibinfo {author}
  {\bibfnamefont {J.}~\bibnamefont {Paglione}},\ }\href {\doibase
  10.1103/PhysRevX.3.011011} {\bibfield  {journal} {\bibinfo  {journal} {Phys.
  Rev. X}\ }\textbf {\bibinfo {volume} {3}},\ \bibinfo {pages} {011011}
  (\bibinfo {year} {2013})}\BibitemShut {NoStop}%
\bibitem [{\citenamefont {Syers}\ \emph {et~al.}(2015)\citenamefont {Syers},
  \citenamefont {Kim}, \citenamefont {Fuhrer},\ and\ \citenamefont
  {Paglione}}]{Syers}%
  \BibitemOpen
  \bibfield  {author} {\bibinfo {author} {\bibfnamefont {P.}~\bibnamefont
  {Syers}}, \bibinfo {author} {\bibfnamefont {D.}~\bibnamefont {Kim}}, \bibinfo
  {author} {\bibfnamefont {M.~S.}\ \bibnamefont {Fuhrer}}, \ and\ \bibinfo
  {author} {\bibfnamefont {J.}~\bibnamefont {Paglione}},\ }\href {\doibase
  10.1103/PhysRevLett.114.096601} {\bibfield  {journal} {\bibinfo  {journal}
  {Phys. Rev. Lett.}\ }\textbf {\bibinfo {volume} {114}},\ \bibinfo {pages}
  {096601} (\bibinfo {year} {2015})}\BibitemShut {NoStop}%
\bibitem [{\citenamefont {Allen}\ \emph {et~al.}(1979)\citenamefont {Allen},
  \citenamefont {Batlogg},\ and\ \citenamefont {Wachter}}]{Allen}%
  \BibitemOpen
  \bibfield  {author} {\bibinfo {author} {\bibfnamefont {J.~W.}\ \bibnamefont
  {Allen}}, \bibinfo {author} {\bibfnamefont {B.}~\bibnamefont {Batlogg}}, \
  and\ \bibinfo {author} {\bibfnamefont {P.}~\bibnamefont {Wachter}},\ }\href
  {\doibase 10.1103/PhysRevB.20.4807} {\bibfield  {journal} {\bibinfo
  {journal} {Phys. Rev. B}\ }\textbf {\bibinfo {volume} {20}},\ \bibinfo
  {pages} {4807} (\bibinfo {year} {1979})}\BibitemShut {NoStop}%
\bibitem [{\citenamefont {Cooley}\ \emph {et~al.}(1995)\citenamefont {Cooley},
  \citenamefont {Aronson}, \citenamefont {Fisk},\ and\ \citenamefont
  {Canfield}}]{Cooley95}%
  \BibitemOpen
  \bibfield  {author} {\bibinfo {author} {\bibfnamefont {J.~C.}\ \bibnamefont
  {Cooley}}, \bibinfo {author} {\bibfnamefont {M.~C.}\ \bibnamefont {Aronson}},
  \bibinfo {author} {\bibfnamefont {Z.}~\bibnamefont {Fisk}}, \ and\ \bibinfo
  {author} {\bibfnamefont {P.~C.}\ \bibnamefont {Canfield}},\ }\href {\doibase
  10.1103/PhysRevLett.74.1629} {\bibfield  {journal} {\bibinfo  {journal}
  {Phys. Rev. Lett.}\ }\textbf {\bibinfo {volume} {74}},\ \bibinfo {pages}
  {1629} (\bibinfo {year} {1995})}\BibitemShut {NoStop}%
\bibitem [{\citenamefont {Cooley}\ \emph {et~al.}(1999)\citenamefont {Cooley},
  \citenamefont {Mielke}, \citenamefont {Hults}, \citenamefont {Goettee},
  \citenamefont {Honold}, \citenamefont {Modler}, \citenamefont {Lacerda},
  \citenamefont {Rickel},\ and\ \citenamefont {Smith}}]{Cooley99}%
  \BibitemOpen
  \bibfield  {author} {\bibinfo {author} {\bibfnamefont {J.~C.}\ \bibnamefont
  {Cooley}}, \bibinfo {author} {\bibfnamefont {C.~H.}\ \bibnamefont {Mielke}},
  \bibinfo {author} {\bibfnamefont {W.~L.}\ \bibnamefont {Hults}}, \bibinfo
  {author} {\bibfnamefont {J.~D.}\ \bibnamefont {Goettee}}, \bibinfo {author}
  {\bibfnamefont {M.~M.}\ \bibnamefont {Honold}}, \bibinfo {author}
  {\bibfnamefont {R.~M.}\ \bibnamefont {Modler}}, \bibinfo {author}
  {\bibfnamefont {A.}~\bibnamefont {Lacerda}}, \bibinfo {author} {\bibfnamefont
  {D.~G.}\ \bibnamefont {Rickel}}, \ and\ \bibinfo {author} {\bibfnamefont
  {J.~L.}\ \bibnamefont {Smith}},\ }\href {\doibase 10.1023/A:1007771030747}
  {\bibfield  {journal} {\bibinfo  {journal} {Journal of Superconductivity}\
  }\textbf {\bibinfo {volume} {12}},\ \bibinfo {pages} {171} (\bibinfo {year}
  {1999})}\BibitemShut {NoStop}%
\bibitem [{\citenamefont {Denlinger}\ \emph {et~al.}(2013)\citenamefont
  {Denlinger}, \citenamefont {Allen}, \citenamefont {Kang}, \citenamefont
  {Sun}, \citenamefont {Min}, \citenamefont {Kim},\ and\ \citenamefont
  {Fisk}}]{Denlinger}%
  \BibitemOpen
  \bibfield  {author} {\bibinfo {author} {\bibfnamefont {J.~D.}\ \bibnamefont
  {Denlinger}}, \bibinfo {author} {\bibfnamefont {J.~W.}\ \bibnamefont
  {Allen}}, \bibinfo {author} {\bibfnamefont {J.-S.}\ \bibnamefont {Kang}},
  \bibinfo {author} {\bibfnamefont {K.}~\bibnamefont {Sun}}, \bibinfo {author}
  {\bibfnamefont {B.-I.}\ \bibnamefont {Min}}, \bibinfo {author} {\bibfnamefont
  {D.-J.}\ \bibnamefont {Kim}}, \ and\ \bibinfo {author} {\bibfnamefont
  {Z.}~\bibnamefont {Fisk}},\ }\href@noop {} {\enquote {\bibinfo {title}
  {{Temperature Dependence of Linked Gap and Surface State Evolution in the
  Mixed Valent Topological Insulator SmB$_{6}$}},}\ } (\bibinfo {year}
  {2013}),\ \Eprint {http://arxiv.org/abs/1312.6637} {arXiv:1312.6637
  [cond-mat.str-el]} \BibitemShut {NoStop}%
\bibitem [{\citenamefont {Tan}\ \emph {et~al.}(2015)\citenamefont {Tan},
  \citenamefont {Hsu}, \citenamefont {Zeng}, \citenamefont {Hatnean},
  \citenamefont {Harrison}, \citenamefont {Zhu}, \citenamefont {Hartstein},
  \citenamefont {Kiourlappou}, \citenamefont {Srivastava}, \citenamefont
  {Johannes}, \citenamefont {Murphy}, \citenamefont {Park}, \citenamefont
  {Balicas}, \citenamefont {Lonzarich}, \citenamefont {Balakrishnan},\ and\
  \citenamefont {Sebastian}}]{Suchitra2015}%
  \BibitemOpen
  \bibfield  {author} {\bibinfo {author} {\bibfnamefont {B.~S.}\ \bibnamefont
  {Tan}}, \bibinfo {author} {\bibfnamefont {Y.-T.}\ \bibnamefont {Hsu}},
  \bibinfo {author} {\bibfnamefont {B.}~\bibnamefont {Zeng}}, \bibinfo {author}
  {\bibfnamefont {M.~C.}\ \bibnamefont {Hatnean}}, \bibinfo {author}
  {\bibfnamefont {N.}~\bibnamefont {Harrison}}, \bibinfo {author}
  {\bibfnamefont {Z.}~\bibnamefont {Zhu}}, \bibinfo {author} {\bibfnamefont
  {M.}~\bibnamefont {Hartstein}}, \bibinfo {author} {\bibfnamefont
  {M.}~\bibnamefont {Kiourlappou}}, \bibinfo {author} {\bibfnamefont
  {A.}~\bibnamefont {Srivastava}}, \bibinfo {author} {\bibfnamefont {M.~D.}\
  \bibnamefont {Johannes}}, \bibinfo {author} {\bibfnamefont {T.~P.}\
  \bibnamefont {Murphy}}, \bibinfo {author} {\bibfnamefont {J.-H.}\
  \bibnamefont {Park}}, \bibinfo {author} {\bibfnamefont {L.}~\bibnamefont
  {Balicas}}, \bibinfo {author} {\bibfnamefont {G.~G.}\ \bibnamefont
  {Lonzarich}}, \bibinfo {author} {\bibfnamefont {G.}~\bibnamefont
  {Balakrishnan}}, \ and\ \bibinfo {author} {\bibfnamefont {S.~E.}\
  \bibnamefont {Sebastian}},\ }\href {\doibase 10.1126/science.aaa7974}
  {\bibfield  {journal} {\bibinfo  {journal} {Science}\ }\textbf {\bibinfo
  {volume} {349}},\ \bibinfo {pages} {287} (\bibinfo {year} {2015})},\ \Eprint
  {http://arxiv.org/abs/http://www.sciencemag.org/content/349/6245/287.full.pdf}
  {http://www.sciencemag.org/content/349/6245/287.full.pdf} \BibitemShut
  {NoStop}%
\bibitem [{\citenamefont {{Li}}\ \emph {et~al.}(2014)\citenamefont {{Li}},
  \citenamefont {{Xiang}}, \citenamefont {{Yu}}, \citenamefont {{Asaba}},
  \citenamefont {{Lawson}}, \citenamefont {{Cai}}, \citenamefont {{Tinsman}},
  \citenamefont {{Berkley}}, \citenamefont {{Wolgast}}, \citenamefont {{Eo}},
  \citenamefont {{Kim}}, \citenamefont {{Kurdak}}, \citenamefont {{Allen}},
  \citenamefont {{Sun}}, \citenamefont {{Chen}}, \citenamefont {{Wang}},
  \citenamefont {{Fisk}},\ and\ \citenamefont {{Li}}}]{Li}%
  \BibitemOpen
  \bibfield  {author} {\bibinfo {author} {\bibfnamefont {G.}~\bibnamefont
  {{Li}}}, \bibinfo {author} {\bibfnamefont {Z.}~\bibnamefont {{Xiang}}},
  \bibinfo {author} {\bibfnamefont {F.}~\bibnamefont {{Yu}}}, \bibinfo {author}
  {\bibfnamefont {T.}~\bibnamefont {{Asaba}}}, \bibinfo {author} {\bibfnamefont
  {B.}~\bibnamefont {{Lawson}}}, \bibinfo {author} {\bibfnamefont
  {P.}~\bibnamefont {{Cai}}}, \bibinfo {author} {\bibfnamefont
  {C.}~\bibnamefont {{Tinsman}}}, \bibinfo {author} {\bibfnamefont
  {A.}~\bibnamefont {{Berkley}}}, \bibinfo {author} {\bibfnamefont
  {S.}~\bibnamefont {{Wolgast}}}, \bibinfo {author} {\bibfnamefont {Y.~S.}\
  \bibnamefont {{Eo}}}, \bibinfo {author} {\bibfnamefont {D.-J.}\ \bibnamefont
  {{Kim}}}, \bibinfo {author} {\bibfnamefont {{\c{C}}.}~\bibnamefont
  {{Kurdak}}}, \bibinfo {author} {\bibfnamefont {J.~W.}\ \bibnamefont
  {{Allen}}}, \bibinfo {author} {\bibfnamefont {K.}~\bibnamefont {{Sun}}},
  \bibinfo {author} {\bibfnamefont {X.~H.}\ \bibnamefont {{Chen}}}, \bibinfo
  {author} {\bibfnamefont {Y.~Y.}\ \bibnamefont {{Wang}}}, \bibinfo {author}
  {\bibfnamefont {Z.}~\bibnamefont {{Fisk}}}, \ and\ \bibinfo {author}
  {\bibfnamefont {L.}~\bibnamefont {{Li}}},\ }\href {\doibase
  10.1126/science.1250366} {\bibfield  {journal} {\bibinfo  {journal}
  {Science}\ }\textbf {\bibinfo {volume} {346}},\ \bibinfo {pages} {1208}
  (\bibinfo {year} {2014})},\ \Eprint
  {http://arxiv.org/abs/http://www.sciencemag.org/content/346/6214/1208.full.pdf}
  {http://www.sciencemag.org/content/346/6214/1208.full.pdf} \BibitemShut
  {NoStop}%
\bibitem [{\citenamefont {Knolle}\ and\ \citenamefont
  {Cooper}(2015)}]{KnolleCooper2015}%
  \BibitemOpen
  \bibfield  {author} {\bibinfo {author} {\bibfnamefont {J.}~\bibnamefont
  {Knolle}}\ and\ \bibinfo {author} {\bibfnamefont {N.~R.}\ \bibnamefont
  {Cooper}},\ }\href {\doibase 10.1103/PhysRevLett.115.146401} {\bibfield
  {journal} {\bibinfo  {journal} {Phys. Rev. Lett.}\ }\textbf {\bibinfo
  {volume} {115}},\ \bibinfo {pages} {146401} (\bibinfo {year}
  {2015})}\BibitemShut {NoStop}%
\bibitem [{\citenamefont {Zhang}\ \emph {et~al.}(2016)\citenamefont {Zhang},
  \citenamefont {Song},\ and\ \citenamefont {Wang}}]{Wang2016}%
  \BibitemOpen
  \bibfield  {author} {\bibinfo {author} {\bibfnamefont {L.}~\bibnamefont
  {Zhang}}, \bibinfo {author} {\bibfnamefont {X.-Y.}\ \bibnamefont {Song}}, \
  and\ \bibinfo {author} {\bibfnamefont {F.}~\bibnamefont {Wang}},\ }\href
  {\doibase 10.1103/PhysRevLett.116.046404} {\bibfield  {journal} {\bibinfo
  {journal} {Phys. Rev. Lett.}\ }\textbf {\bibinfo {volume} {116}},\ \bibinfo
  {pages} {046404} (\bibinfo {year} {2016})}\BibitemShut {NoStop}%
\bibitem [{\citenamefont {Erten}\ \emph {et~al.}(2016)\citenamefont {Erten},
  \citenamefont {Ghaemi},\ and\ \citenamefont {Coleman}}]{Erten2016}%
  \BibitemOpen
  \bibfield  {author} {\bibinfo {author} {\bibfnamefont {O.}~\bibnamefont
  {Erten}}, \bibinfo {author} {\bibfnamefont {P.}~\bibnamefont {Ghaemi}}, \
  and\ \bibinfo {author} {\bibfnamefont {P.}~\bibnamefont {Coleman}},\ }\href
  {\doibase 10.1103/PhysRevLett.116.046403} {\bibfield  {journal} {\bibinfo
  {journal} {Phys. Rev. Lett.}\ }\textbf {\bibinfo {volume} {116}},\ \bibinfo
  {pages} {046403} (\bibinfo {year} {2016})}\BibitemShut {NoStop}%
\bibitem [{\citenamefont {Thomson}\ and\ \citenamefont
  {Sachdev}(2016)}]{ThomsonSachdev2016}%
  \BibitemOpen
  \bibfield  {author} {\bibinfo {author} {\bibfnamefont {A.}~\bibnamefont
  {Thomson}}\ and\ \bibinfo {author} {\bibfnamefont {S.}~\bibnamefont
  {Sachdev}},\ }\href {\doibase 10.1103/PhysRevB.93.125103} {\bibfield
  {journal} {\bibinfo  {journal} {Phys. Rev. B}\ }\textbf {\bibinfo {volume}
  {93}},\ \bibinfo {pages} {125103} (\bibinfo {year} {2016})}\BibitemShut
  {NoStop}%
\bibitem [{\citenamefont {Pal}\ \emph {et~al.}(2016)\citenamefont {Pal},
  \citenamefont {Pi\'{e}chon}, \citenamefont {Fuchs}, \citenamefont {Goerbig},\
  and\ \citenamefont {Montambaux}}]{Montambaux2016}%
  \BibitemOpen
  \bibfield  {author} {\bibinfo {author} {\bibfnamefont {H.~K.}\ \bibnamefont
  {Pal}}, \bibinfo {author} {\bibfnamefont {F.}~\bibnamefont {Pi\'{e}chon}},
  \bibinfo {author} {\bibfnamefont {J.-N.}\ \bibnamefont {Fuchs}}, \bibinfo
  {author} {\bibfnamefont {M.}~\bibnamefont {Goerbig}}, \ and\ \bibinfo
  {author} {\bibfnamefont {G.}~\bibnamefont {Montambaux}},\ }\href@noop {}
  {\enquote {\bibinfo {title} {{Quantum Oscillations in Gapped Systems}},}\ }
  (\bibinfo {year} {2016}),\ \Eprint {http://arxiv.org/abs/1604.01688}
  {arXiv:1604.01688} \BibitemShut {NoStop}%
\bibitem [{\citenamefont {Menth}\ \emph {et~al.}(1969)\citenamefont {Menth},
  \citenamefont {Buehler},\ and\ \citenamefont {Geballe}}]{Menth}%
  \BibitemOpen
  \bibfield  {author} {\bibinfo {author} {\bibfnamefont {A.}~\bibnamefont
  {Menth}}, \bibinfo {author} {\bibfnamefont {E.}~\bibnamefont {Buehler}}, \
  and\ \bibinfo {author} {\bibfnamefont {T.~H.}\ \bibnamefont {Geballe}},\
  }\href {\doibase 10.1103/PhysRevLett.22.295} {\bibfield  {journal} {\bibinfo
  {journal} {Phys. Rev. Lett.}\ }\textbf {\bibinfo {volume} {22}},\ \bibinfo
  {pages} {295} (\bibinfo {year} {1969})}\BibitemShut {NoStop}%
\bibitem [{\citenamefont {Gorshunov}\ \emph {et~al.}(1999)\citenamefont
  {Gorshunov}, \citenamefont {Sluchanko}, \citenamefont {Volkov}, \citenamefont
  {Dressel}, \citenamefont {Knebel}, \citenamefont {Loidl},\ and\ \citenamefont
  {Kunii}}]{Dressel1999}%
  \BibitemOpen
  \bibfield  {author} {\bibinfo {author} {\bibfnamefont {B.}~\bibnamefont
  {Gorshunov}}, \bibinfo {author} {\bibfnamefont {N.}~\bibnamefont
  {Sluchanko}}, \bibinfo {author} {\bibfnamefont {A.}~\bibnamefont {Volkov}},
  \bibinfo {author} {\bibfnamefont {M.}~\bibnamefont {Dressel}}, \bibinfo
  {author} {\bibfnamefont {G.}~\bibnamefont {Knebel}}, \bibinfo {author}
  {\bibfnamefont {A.}~\bibnamefont {Loidl}}, \ and\ \bibinfo {author}
  {\bibfnamefont {S.}~\bibnamefont {Kunii}},\ }\href {\doibase
  10.1103/PhysRevB.59.1808} {\bibfield  {journal} {\bibinfo  {journal} {Phys.
  Rev. B}\ }\textbf {\bibinfo {volume} {59}},\ \bibinfo {pages} {1808}
  (\bibinfo {year} {1999})}\BibitemShut {NoStop}%
\bibitem [{\citenamefont {Flachbart}\ \emph
  {et~al.}(2001{\natexlab{a}})\citenamefont {Flachbart}, \citenamefont
  {Gab{\'{a}}ni}, \citenamefont {Konovalova}, \citenamefont {Paderno},\ and\
  \citenamefont {Pavl{\'{i}}k}}]{Flachbart}%
  \BibitemOpen
  \bibfield  {author} {\bibinfo {author} {\bibfnamefont {K.}~\bibnamefont
  {Flachbart}}, \bibinfo {author} {\bibfnamefont {S.}~\bibnamefont
  {Gab{\'{a}}ni}}, \bibinfo {author} {\bibfnamefont {E.}~\bibnamefont
  {Konovalova}}, \bibinfo {author} {\bibfnamefont {Y.}~\bibnamefont {Paderno}},
  \ and\ \bibinfo {author} {\bibfnamefont {V.}~\bibnamefont {Pavl{\'{i}}k}},\
  }\href {\doibase http://dx.doi.org/10.1016/S0921-4526(00)00552-4} {\bibfield
  {journal} {\bibinfo  {journal} {Physica B: Condensed Matter}\ }\textbf
  {\bibinfo {volume} {293}},\ \bibinfo {pages} {417} (\bibinfo {year}
  {2001}{\natexlab{a}})}\BibitemShut {NoStop}%
\bibitem [{\citenamefont {Flachbart}\ \emph
  {et~al.}(2001{\natexlab{b}})\citenamefont {Flachbart}, \citenamefont {Gloos},
  \citenamefont {Konovalova}, \citenamefont {Paderno}, \citenamefont
  {Reiffers}, \citenamefont {Samuely},\ and\ \citenamefont
  {\ifmmode~\check{S}\else \v{S}\fi{}vec}}]{FlachbartSpect2001}%
  \BibitemOpen
  \bibfield  {author} {\bibinfo {author} {\bibfnamefont {K.}~\bibnamefont
  {Flachbart}}, \bibinfo {author} {\bibfnamefont {K.}~\bibnamefont {Gloos}},
  \bibinfo {author} {\bibfnamefont {E.}~\bibnamefont {Konovalova}}, \bibinfo
  {author} {\bibfnamefont {Y.}~\bibnamefont {Paderno}}, \bibinfo {author}
  {\bibfnamefont {M.}~\bibnamefont {Reiffers}}, \bibinfo {author}
  {\bibfnamefont {P.}~\bibnamefont {Samuely}}, \ and\ \bibinfo {author}
  {\bibfnamefont {P.}~\bibnamefont {\ifmmode~\check{S}\else \v{S}\fi{}vec}},\
  }\href {\doibase 10.1103/PhysRevB.64.085104} {\bibfield  {journal} {\bibinfo
  {journal} {Phys. Rev. B}\ }\textbf {\bibinfo {volume} {64}},\ \bibinfo
  {pages} {085104} (\bibinfo {year} {2001}{\natexlab{b}})}\BibitemShut
  {NoStop}%
\bibitem [{\citenamefont {Gab\'{a}ni}\ \emph {et~al.}(2001)\citenamefont
  {Gab\'{a}ni}, \citenamefont {Flachbart}, \citenamefont {Konovalova},
  \citenamefont {Orend\'{a}\v{c}}, \citenamefont {Paderno}, \citenamefont
  {Pavl\'{i}k},\ and\ \citenamefont {\v{S}ebek}}]{Gabani2001}%
  \BibitemOpen
  \bibfield  {author} {\bibinfo {author} {\bibfnamefont {S.}~\bibnamefont
  {Gab\'{a}ni}}, \bibinfo {author} {\bibfnamefont {K.}~\bibnamefont
  {Flachbart}}, \bibinfo {author} {\bibfnamefont {E.}~\bibnamefont
  {Konovalova}}, \bibinfo {author} {\bibfnamefont {M.}~\bibnamefont
  {Orend\'{a}\v{c}}}, \bibinfo {author} {\bibfnamefont {Y.}~\bibnamefont
  {Paderno}}, \bibinfo {author} {\bibfnamefont {V.}~\bibnamefont {Pavl\'{i}k}},
  \ and\ \bibinfo {author} {\bibfnamefont {J.}~\bibnamefont {\v{S}ebek}},\
  }\href {\doibase http://dx.doi.org/10.1016/S0038-1098(01)00004-7} {\bibfield
  {journal} {\bibinfo  {journal} {Solid State Communications}\ }\textbf
  {\bibinfo {volume} {117}},\ \bibinfo {pages} {641} (\bibinfo {year}
  {2001})}\BibitemShut {NoStop}%
\bibitem [{\citenamefont {Neupane}\ \emph {et~al.}(2013)\citenamefont
  {Neupane}, \citenamefont {Alidoust}, \citenamefont {Xu}, \citenamefont
  {Kondo}, \citenamefont {Ishida}, \citenamefont {Kim}, \citenamefont {Liu},
  \citenamefont {Belopolski}, \citenamefont {Jo}, \citenamefont {Chang},
  \citenamefont {Jeng}, \citenamefont {Durakiewicz}, \citenamefont {Balicas},
  \citenamefont {Lin}, \citenamefont {Bansil}, \citenamefont {Shin},
  \citenamefont {Fisk},\ and\ \citenamefont {Hasan}}]{Neupane}%
  \BibitemOpen
  \bibfield  {author} {\bibinfo {author} {\bibfnamefont {M.}~\bibnamefont
  {Neupane}}, \bibinfo {author} {\bibfnamefont {N.}~\bibnamefont {Alidoust}},
  \bibinfo {author} {\bibfnamefont {S.-Y.}\ \bibnamefont {Xu}}, \bibinfo
  {author} {\bibfnamefont {T.}~\bibnamefont {Kondo}}, \bibinfo {author}
  {\bibfnamefont {Y.}~\bibnamefont {Ishida}}, \bibinfo {author} {\bibfnamefont
  {D.~J.}\ \bibnamefont {Kim}}, \bibinfo {author} {\bibfnamefont
  {C.}~\bibnamefont {Liu}}, \bibinfo {author} {\bibfnamefont {I.}~\bibnamefont
  {Belopolski}}, \bibinfo {author} {\bibfnamefont {Y.~J.}\ \bibnamefont {Jo}},
  \bibinfo {author} {\bibfnamefont {T.-R.}\ \bibnamefont {Chang}}, \bibinfo
  {author} {\bibfnamefont {H.-T.}\ \bibnamefont {Jeng}}, \bibinfo {author}
  {\bibfnamefont {T.}~\bibnamefont {Durakiewicz}}, \bibinfo {author}
  {\bibfnamefont {L.}~\bibnamefont {Balicas}}, \bibinfo {author} {\bibfnamefont
  {H.}~\bibnamefont {Lin}}, \bibinfo {author} {\bibfnamefont {A.}~\bibnamefont
  {Bansil}}, \bibinfo {author} {\bibfnamefont {S.}~\bibnamefont {Shin}},
  \bibinfo {author} {\bibfnamefont {Z.}~\bibnamefont {Fisk}}, \ and\ \bibinfo
  {author} {\bibfnamefont {M.~Z.}\ \bibnamefont {Hasan}},\ }\href {\doibase
  doi:10.1038/ncomms3991} {\bibfield  {journal} {\bibinfo  {journal} {Nat.
  Commun.}\ }\textbf {\bibinfo {volume} {4}},\ \bibinfo {pages} {2991}
  (\bibinfo {year} {2013})}\BibitemShut {NoStop}%
\bibitem [{\citenamefont {Miyazaki}\ \emph {et~al.}(2012)\citenamefont
  {Miyazaki}, \citenamefont {Hajiri}, \citenamefont {Ito}, \citenamefont
  {Kunii},\ and\ \citenamefont {Kimura}}]{Miyazaki}%
  \BibitemOpen
  \bibfield  {author} {\bibinfo {author} {\bibfnamefont {H.}~\bibnamefont
  {Miyazaki}}, \bibinfo {author} {\bibfnamefont {T.}~\bibnamefont {Hajiri}},
  \bibinfo {author} {\bibfnamefont {T.}~\bibnamefont {Ito}}, \bibinfo {author}
  {\bibfnamefont {S.}~\bibnamefont {Kunii}}, \ and\ \bibinfo {author}
  {\bibfnamefont {S.~I.}\ \bibnamefont {Kimura}},\ }\href {\doibase
  10.1103/PhysRevB.86.075105} {\bibfield  {journal} {\bibinfo  {journal} {Phys.
  Rev. B}\ }\textbf {\bibinfo {volume} {86}},\ \bibinfo {pages} {075105}
  (\bibinfo {year} {2012})}\BibitemShut {NoStop}%
\bibitem [{\citenamefont {Jiang}\ \emph {et~al.}(2013)\citenamefont {Jiang},
  \citenamefont {Li}, \citenamefont {Zhang}, \citenamefont {Sun}, \citenamefont
  {Chen}, \citenamefont {Ye}, \citenamefont {Xu}, \citenamefont {Ge},
  \citenamefont {Tan}, \citenamefont {Niu}, \citenamefont {Xia}, \citenamefont
  {Xie}, \citenamefont {Li}, \citenamefont {Chen}, \citenamefont {Wen},\ and\
  \citenamefont {Feng}}]{Jiang}%
  \BibitemOpen
  \bibfield  {author} {\bibinfo {author} {\bibfnamefont {J.}~\bibnamefont
  {Jiang}}, \bibinfo {author} {\bibfnamefont {S.}~\bibnamefont {Li}}, \bibinfo
  {author} {\bibfnamefont {T.}~\bibnamefont {Zhang}}, \bibinfo {author}
  {\bibfnamefont {Z.}~\bibnamefont {Sun}}, \bibinfo {author} {\bibfnamefont
  {F.}~\bibnamefont {Chen}}, \bibinfo {author} {\bibfnamefont {Z.~R.}\
  \bibnamefont {Ye}}, \bibinfo {author} {\bibfnamefont {M.}~\bibnamefont {Xu}},
  \bibinfo {author} {\bibfnamefont {Q.~Q.}\ \bibnamefont {Ge}}, \bibinfo
  {author} {\bibfnamefont {S.~Y.}\ \bibnamefont {Tan}}, \bibinfo {author}
  {\bibfnamefont {X.~H.}\ \bibnamefont {Niu}}, \bibinfo {author} {\bibfnamefont
  {M.}~\bibnamefont {Xia}}, \bibinfo {author} {\bibfnamefont {B.~P.}\
  \bibnamefont {Xie}}, \bibinfo {author} {\bibfnamefont {Y.~F.}\ \bibnamefont
  {Li}}, \bibinfo {author} {\bibfnamefont {X.~H.}\ \bibnamefont {Chen}},
  \bibinfo {author} {\bibfnamefont {H.~H.}\ \bibnamefont {Wen}}, \ and\
  \bibinfo {author} {\bibfnamefont {D.~L.}\ \bibnamefont {Feng}},\ }\href
  {\doibase 10.1038/ncomms4010} {\bibfield  {journal} {\bibinfo  {journal}
  {Nat. Commun.}\ }\textbf {\bibinfo {volume} {4}},\ \bibinfo {pages} {3010}
  (\bibinfo {year} {2013})}\BibitemShut {NoStop}%
\bibitem [{\citenamefont {Xu}\ \emph {et~al.}(2014)\citenamefont {Xu},
  \citenamefont {Biswas}, \citenamefont {Dil}, \citenamefont {Dhaka},
  \citenamefont {Landolt}, \citenamefont {Muff}, \citenamefont {Matt},
  \citenamefont {Shi}, \citenamefont {Plumb}, \citenamefont {Radovi{\'c}},
  \citenamefont {Pomjakushina}, \citenamefont {Conder}, \citenamefont {Amato},
  \citenamefont {Borisenko}, \citenamefont {Yu}, \citenamefont {Weng},
  \citenamefont {Fang}, \citenamefont {Dai}, \citenamefont {Mesot},
  \citenamefont {Ding},\ and\ \citenamefont {Shi}}]{XuSARPES}%
  \BibitemOpen
  \bibfield  {author} {\bibinfo {author} {\bibfnamefont {N.}~\bibnamefont
  {Xu}}, \bibinfo {author} {\bibfnamefont {P.~K.}\ \bibnamefont {Biswas}},
  \bibinfo {author} {\bibfnamefont {J.~H.}\ \bibnamefont {Dil}}, \bibinfo
  {author} {\bibfnamefont {R.~S.}\ \bibnamefont {Dhaka}}, \bibinfo {author}
  {\bibfnamefont {G.}~\bibnamefont {Landolt}}, \bibinfo {author} {\bibfnamefont
  {S.}~\bibnamefont {Muff}}, \bibinfo {author} {\bibfnamefont {C.~E.}\
  \bibnamefont {Matt}}, \bibinfo {author} {\bibfnamefont {X.}~\bibnamefont
  {Shi}}, \bibinfo {author} {\bibfnamefont {N.~C.}\ \bibnamefont {Plumb}},
  \bibinfo {author} {\bibfnamefont {M.}~\bibnamefont {Radovi{\'c}}}, \bibinfo
  {author} {\bibfnamefont {E.}~\bibnamefont {Pomjakushina}}, \bibinfo {author}
  {\bibfnamefont {K.}~\bibnamefont {Conder}}, \bibinfo {author} {\bibfnamefont
  {A.}~\bibnamefont {Amato}}, \bibinfo {author} {\bibfnamefont {S.~V.}\
  \bibnamefont {Borisenko}}, \bibinfo {author} {\bibfnamefont {R.}~\bibnamefont
  {Yu}}, \bibinfo {author} {\bibfnamefont {H.-M.}\ \bibnamefont {Weng}},
  \bibinfo {author} {\bibfnamefont {Z.}~\bibnamefont {Fang}}, \bibinfo {author}
  {\bibfnamefont {X.}~\bibnamefont {Dai}}, \bibinfo {author} {\bibfnamefont
  {J.}~\bibnamefont {Mesot}}, \bibinfo {author} {\bibfnamefont
  {H.}~\bibnamefont {Ding}}, \ and\ \bibinfo {author} {\bibfnamefont
  {M.}~\bibnamefont {Shi}},\ }\href {\doibase doi:10.1038/ncomms5566}
  {\bibfield  {journal} {\bibinfo  {journal} {Nat. Commun.}\ }\textbf {\bibinfo
  {volume} {5}},\ \bibinfo {pages} {4566} (\bibinfo {year} {2014})}\BibitemShut
  {NoStop}%
\bibitem [{\citenamefont {Xu}\ \emph {et~al.}(2013)\citenamefont {Xu},
  \citenamefont {Shi}, \citenamefont {Biswas}, \citenamefont {Matt},
  \citenamefont {Dhaka}, \citenamefont {Huang}, \citenamefont {Plumb},
  \citenamefont {Radovi\ifmmode~\acute{c}\else \'{c}\fi{}}, \citenamefont
  {Dil}, \citenamefont {Pomjakushina}, \citenamefont {Conder}, \citenamefont
  {Amato}, \citenamefont {Salman}, \citenamefont {Paul}, \citenamefont {Mesot},
  \citenamefont {Ding},\ and\ \citenamefont {Shi}}]{NXu}%
  \BibitemOpen
  \bibfield  {author} {\bibinfo {author} {\bibfnamefont {N.}~\bibnamefont
  {Xu}}, \bibinfo {author} {\bibfnamefont {X.}~\bibnamefont {Shi}}, \bibinfo
  {author} {\bibfnamefont {P.~K.}\ \bibnamefont {Biswas}}, \bibinfo {author}
  {\bibfnamefont {C.~E.}\ \bibnamefont {Matt}}, \bibinfo {author}
  {\bibfnamefont {R.~S.}\ \bibnamefont {Dhaka}}, \bibinfo {author}
  {\bibfnamefont {Y.}~\bibnamefont {Huang}}, \bibinfo {author} {\bibfnamefont
  {N.~C.}\ \bibnamefont {Plumb}}, \bibinfo {author} {\bibfnamefont
  {M.}~\bibnamefont {Radovi\ifmmode~\acute{c}\else \'{c}\fi{}}}, \bibinfo
  {author} {\bibfnamefont {J.~H.}\ \bibnamefont {Dil}}, \bibinfo {author}
  {\bibfnamefont {E.}~\bibnamefont {Pomjakushina}}, \bibinfo {author}
  {\bibfnamefont {K.}~\bibnamefont {Conder}}, \bibinfo {author} {\bibfnamefont
  {A.}~\bibnamefont {Amato}}, \bibinfo {author} {\bibfnamefont
  {Z.}~\bibnamefont {Salman}}, \bibinfo {author} {\bibfnamefont {D.~M.}\
  \bibnamefont {Paul}}, \bibinfo {author} {\bibfnamefont {J.}~\bibnamefont
  {Mesot}}, \bibinfo {author} {\bibfnamefont {H.}~\bibnamefont {Ding}}, \ and\
  \bibinfo {author} {\bibfnamefont {M.}~\bibnamefont {Shi}},\ }\href {\doibase
  10.1103/PhysRevB.88.121102} {\bibfield  {journal} {\bibinfo  {journal} {Phys.
  Rev. B(R)}\ }\textbf {\bibinfo {volume} {88}},\ \bibinfo {pages} {121102}
  (\bibinfo {year} {2013})}\BibitemShut {NoStop}%
\bibitem [{\citenamefont {Hlawenka}\ \emph {et~al.}(2015)\citenamefont
  {Hlawenka}, \citenamefont {Siemensmeyer}, \citenamefont {Weschke},
  \citenamefont {Varykhalov}, \citenamefont {S{\'{a}}nchez-Barriga},
  \citenamefont {Shitsevalova}, \citenamefont {Dukhnenko}, \citenamefont
  {Filipov}, \citenamefont {Gab{\'{a}}ni}, \citenamefont {Flachbart},
  \citenamefont {Rader},\ and\ \citenamefont {Rienks}}]{Rader}%
  \BibitemOpen
  \bibfield  {author} {\bibinfo {author} {\bibfnamefont {P.}~\bibnamefont
  {Hlawenka}}, \bibinfo {author} {\bibfnamefont {K.}~\bibnamefont
  {Siemensmeyer}}, \bibinfo {author} {\bibfnamefont {E.}~\bibnamefont
  {Weschke}}, \bibinfo {author} {\bibfnamefont {A.}~\bibnamefont {Varykhalov}},
  \bibinfo {author} {\bibfnamefont {J.}~\bibnamefont {S{\'{a}}nchez-Barriga}},
  \bibinfo {author} {\bibfnamefont {N.~Y.}\ \bibnamefont {Shitsevalova}},
  \bibinfo {author} {\bibfnamefont {A.~V.}\ \bibnamefont {Dukhnenko}}, \bibinfo
  {author} {\bibfnamefont {V.~B.}\ \bibnamefont {Filipov}}, \bibinfo {author}
  {\bibfnamefont {S.}~\bibnamefont {Gab{\'{a}}ni}}, \bibinfo {author}
  {\bibfnamefont {K.}~\bibnamefont {Flachbart}}, \bibinfo {author}
  {\bibfnamefont {O.}~\bibnamefont {Rader}}, \ and\ \bibinfo {author}
  {\bibfnamefont {E.~D.~L.}\ \bibnamefont {Rienks}},\ }\href@noop {} {\enquote
  {\bibinfo {title} {{Samarium hexaboride: A trivial surface conductor}},}\ }
  (\bibinfo {year} {2015}),\ \Eprint {http://arxiv.org/abs/1502.01542}
  {arXiv:1502.01542 [cond-mat.str-el]} \BibitemShut {NoStop}%
\bibitem [{\citenamefont {Rakoski}\ \emph {et~al.}(2016)\citenamefont
  {Rakoski}, \citenamefont {Eo}, \citenamefont {Sun},\ and\ \citenamefont
  {Kurdak}}]{AlexaDepletionWidth}%
  \BibitemOpen
  \bibfield  {author} {\bibinfo {author} {\bibfnamefont {A.}~\bibnamefont
  {Rakoski}}, \bibinfo {author} {\bibfnamefont {Y.~S.}\ \bibnamefont {Eo}},
  \bibinfo {author} {\bibfnamefont {K.}~\bibnamefont {Sun}}, \ and\ \bibinfo
  {author} {\bibfnamefont {{\c{C}}.}~\bibnamefont {Kurdak}},\ }\href@noop {}
  {\enquote {\bibinfo {title} {Understanding the low-temperature gap in
  samarium hexaboride without relying on in-gap states},}\ } (\bibinfo {year}
  {2016}),\ \bibinfo {note} {unpublished}\BibitemShut {NoStop}%
\bibitem [{\citenamefont {Nozawa}\ \emph {et~al.}(2002)\citenamefont {Nozawa},
  \citenamefont {Tsukamoto}, \citenamefont {Kanai}, \citenamefont {Haruna},
  \citenamefont {Shin},\ and\ \citenamefont {Kunii}}]{Nozawa2002}%
  \BibitemOpen
  \bibfield  {author} {\bibinfo {author} {\bibfnamefont {S.}~\bibnamefont
  {Nozawa}}, \bibinfo {author} {\bibfnamefont {T.}~\bibnamefont {Tsukamoto}},
  \bibinfo {author} {\bibfnamefont {K.}~\bibnamefont {Kanai}}, \bibinfo
  {author} {\bibfnamefont {T.}~\bibnamefont {Haruna}}, \bibinfo {author}
  {\bibfnamefont {S.}~\bibnamefont {Shin}}, \ and\ \bibinfo {author}
  {\bibfnamefont {S.}~\bibnamefont {Kunii}},\ }\href {\doibase
  http://dx.doi.org/10.1016/S0022-3697(02)00020-3} {\bibfield  {journal}
  {\bibinfo  {journal} {Journal of Physics and Chemistry of Solids}\ }\textbf
  {\bibinfo {volume} {63}},\ \bibinfo {pages} {1223} (\bibinfo {year}
  {2002})},\ \bibinfo {note} {proceedings of the 8th {ISSP} International
  Symposium}\BibitemShut {NoStop}%
\bibitem [{\citenamefont {Flachbart}\ \emph {et~al.}(2009)\citenamefont
  {Flachbart}, \citenamefont {Bartkowiak}, \citenamefont {Demishev},
  \citenamefont {Gabani}, \citenamefont {Glushkov}, \citenamefont
  {Herrmannsdorfer}, \citenamefont {Moshchalkov}, \citenamefont
  {Shitsevalova},\ and\ \citenamefont {Sluchanko}}]{Flachbart2009}%
  \BibitemOpen
  \bibfield  {author} {\bibinfo {author} {\bibfnamefont {K.}~\bibnamefont
  {Flachbart}}, \bibinfo {author} {\bibfnamefont {M.}~\bibnamefont
  {Bartkowiak}}, \bibinfo {author} {\bibfnamefont {S.}~\bibnamefont
  {Demishev}}, \bibinfo {author} {\bibfnamefont {S.}~\bibnamefont {Gabani}},
  \bibinfo {author} {\bibfnamefont {V.}~\bibnamefont {Glushkov}}, \bibinfo
  {author} {\bibfnamefont {T.}~\bibnamefont {Herrmannsdorfer}}, \bibinfo
  {author} {\bibfnamefont {V.}~\bibnamefont {Moshchalkov}}, \bibinfo {author}
  {\bibfnamefont {N.}~\bibnamefont {Shitsevalova}}, \ and\ \bibinfo {author}
  {\bibfnamefont {N.}~\bibnamefont {Sluchanko}},\ }\href {\doibase
  http://dx.doi.org/10.1016/j.physb.2009.07.017} {\bibfield  {journal}
  {\bibinfo  {journal} {Physica B: Condensed Matter}\ }\textbf {\bibinfo
  {volume} {404}},\ \bibinfo {pages} {2985} (\bibinfo {year} {2009})},\
  \bibinfo {note} {proceedings of the International Conference on Strongly
  Correlated Electron Systems}\BibitemShut {NoStop}%
\bibitem [{\citenamefont {Wolgast}\ \emph {et~al.}(2015)\citenamefont
  {Wolgast}, \citenamefont {Eo}, \citenamefont {\"Ozt\"urk}, \citenamefont
  {Li}, \citenamefont {Xiang}, \citenamefont {Tinsman}, \citenamefont {Asaba},
  \citenamefont {Lawson}, \citenamefont {Yu}, \citenamefont {Allen},
  \citenamefont {Sun}, \citenamefont {Li}, \citenamefont {Kurdak},
  \citenamefont {Kim},\ and\ \citenamefont {Fisk}}]{WolgastMR}%
  \BibitemOpen
  \bibfield  {author} {\bibinfo {author} {\bibfnamefont {S.}~\bibnamefont
  {Wolgast}}, \bibinfo {author} {\bibfnamefont {Y.~S.}\ \bibnamefont {Eo}},
  \bibinfo {author} {\bibfnamefont {T.}~\bibnamefont {\"Ozt\"urk}}, \bibinfo
  {author} {\bibfnamefont {G.}~\bibnamefont {Li}}, \bibinfo {author}
  {\bibfnamefont {Z.}~\bibnamefont {Xiang}}, \bibinfo {author} {\bibfnamefont
  {C.}~\bibnamefont {Tinsman}}, \bibinfo {author} {\bibfnamefont
  {T.}~\bibnamefont {Asaba}}, \bibinfo {author} {\bibfnamefont
  {B.}~\bibnamefont {Lawson}}, \bibinfo {author} {\bibfnamefont
  {F.}~\bibnamefont {Yu}}, \bibinfo {author} {\bibfnamefont {J.~W.}\
  \bibnamefont {Allen}}, \bibinfo {author} {\bibfnamefont {K.}~\bibnamefont
  {Sun}}, \bibinfo {author} {\bibfnamefont {L.}~\bibnamefont {Li}}, \bibinfo
  {author} {\bibfnamefont {i.~m.~c.}\ \bibnamefont {Kurdak}}, \bibinfo {author}
  {\bibfnamefont {D.-J.}\ \bibnamefont {Kim}}, \ and\ \bibinfo {author}
  {\bibfnamefont {Z.}~\bibnamefont {Fisk}},\ }\href {\doibase
  10.1103/PhysRevB.92.115110} {\bibfield  {journal} {\bibinfo  {journal} {Phys.
  Rev. B}\ }\textbf {\bibinfo {volume} {92}},\ \bibinfo {pages} {115110}
  (\bibinfo {year} {2015})}\BibitemShut {NoStop}%
\bibitem [{\citenamefont {Sugiyama}\ \emph {et~al.}(1988)\citenamefont
  {Sugiyama}, \citenamefont {Iga}, \citenamefont {Kasaya}, \citenamefont
  {Kasuya},\ and\ \citenamefont {Date}}]{Sugiyama1988}%
  \BibitemOpen
  \bibfield  {author} {\bibinfo {author} {\bibfnamefont {K.}~\bibnamefont
  {Sugiyama}}, \bibinfo {author} {\bibfnamefont {F.}~\bibnamefont {Iga}},
  \bibinfo {author} {\bibfnamefont {M.}~\bibnamefont {Kasaya}}, \bibinfo
  {author} {\bibfnamefont {T.}~\bibnamefont {Kasuya}}, \ and\ \bibinfo {author}
  {\bibfnamefont {M.}~\bibnamefont {Date}},\ }\href {\doibase
  10.1143/JPSJ.57.3946} {\bibfield  {journal} {\bibinfo  {journal} {Journal of
  the Physical Society of Japan}\ }\textbf {\bibinfo {volume} {57}},\ \bibinfo
  {pages} {3946} (\bibinfo {year} {1988})},\ \Eprint
  {http://arxiv.org/abs/http://dx.doi.org/10.1143/JPSJ.57.3946}
  {http://dx.doi.org/10.1143/JPSJ.57.3946} \BibitemShut {NoStop}%
\bibitem [{\citenamefont {Thomas}\ \emph {et~al.}(2013)\citenamefont {Thomas},
  \citenamefont {Kim}, \citenamefont {Chung}, \citenamefont {Grant},
  \citenamefont {Fisk},\ and\ \citenamefont {Xia}}]{Thomas}%
  \BibitemOpen
  \bibfield  {author} {\bibinfo {author} {\bibfnamefont {S.}~\bibnamefont
  {Thomas}}, \bibinfo {author} {\bibfnamefont {D.~J.}\ \bibnamefont {Kim}},
  \bibinfo {author} {\bibfnamefont {S.~B.}\ \bibnamefont {Chung}}, \bibinfo
  {author} {\bibfnamefont {T.}~\bibnamefont {Grant}}, \bibinfo {author}
  {\bibfnamefont {Z.}~\bibnamefont {Fisk}}, \ and\ \bibinfo {author}
  {\bibfnamefont {J.}~\bibnamefont {Xia}},\ }\href@noop {} {\enquote {\bibinfo
  {title} {{Weak Antilocalization and Linear Magnetoresistance in the Surface
  State of SmB$_6$}},}\ } (\bibinfo {year} {2013}),\ \Eprint
  {http://arxiv.org/abs/1307.4133} {arXiv:1307.4133 [cond-mat.str-el]}
  \BibitemShut {NoStop}%
\bibitem [{\citenamefont {Nakajima}\ \emph {et~al.}(2013)\citenamefont
  {Nakajima}, \citenamefont {Syers}, \citenamefont {Wang}, \citenamefont
  {Wang},\ and\ \citenamefont {Paglione}}]{Nakajima}%
  \BibitemOpen
  \bibfield  {author} {\bibinfo {author} {\bibfnamefont {Y.}~\bibnamefont
  {Nakajima}}, \bibinfo {author} {\bibfnamefont {P.~S.}\ \bibnamefont {Syers}},
  \bibinfo {author} {\bibfnamefont {X.}~\bibnamefont {Wang}}, \bibinfo {author}
  {\bibfnamefont {R.}~\bibnamefont {Wang}}, \ and\ \bibinfo {author}
  {\bibfnamefont {J.}~\bibnamefont {Paglione}},\ }\href@noop {} {\enquote
  {\bibinfo {title} {{One-Dimensional Edge State Transport in a Topological
  Kondo Insulator}},}\ } (\bibinfo {year} {2013}),\ \Eprint
  {http://arxiv.org/abs/1312.6132} {arXiv:1312.6132 [cond-mat.str-el]}
  \BibitemShut {NoStop}%
\bibitem [{\citenamefont {Chen}\ \emph {et~al.}(2015)\citenamefont {Chen},
  \citenamefont {Shang}, \citenamefont {Jin}, \citenamefont {Zhao},
  \citenamefont {Wu}, \citenamefont {Xiang}, \citenamefont {Xia}, \citenamefont
  {Wang}, \citenamefont {Luo}, \citenamefont {Wu},\ and\ \citenamefont
  {Chen}}]{Chen2015}%
  \BibitemOpen
  \bibfield  {author} {\bibinfo {author} {\bibfnamefont {F.}~\bibnamefont
  {Chen}}, \bibinfo {author} {\bibfnamefont {C.}~\bibnamefont {Shang}},
  \bibinfo {author} {\bibfnamefont {Z.}~\bibnamefont {Jin}}, \bibinfo {author}
  {\bibfnamefont {D.}~\bibnamefont {Zhao}}, \bibinfo {author} {\bibfnamefont
  {Y.~P.}\ \bibnamefont {Wu}}, \bibinfo {author} {\bibfnamefont {Z.~J.}\
  \bibnamefont {Xiang}}, \bibinfo {author} {\bibfnamefont {Z.~C.}\ \bibnamefont
  {Xia}}, \bibinfo {author} {\bibfnamefont {A.~F.}\ \bibnamefont {Wang}},
  \bibinfo {author} {\bibfnamefont {X.~G.}\ \bibnamefont {Luo}}, \bibinfo
  {author} {\bibfnamefont {T.}~\bibnamefont {Wu}}, \ and\ \bibinfo {author}
  {\bibfnamefont {X.~H.}\ \bibnamefont {Chen}},\ }\href {\doibase
  10.1103/PhysRevB.91.205133} {\bibfield  {journal} {\bibinfo  {journal} {Phys.
  Rev. B}\ }\textbf {\bibinfo {volume} {91}},\ \bibinfo {pages} {205133}
  (\bibinfo {year} {2015})}\BibitemShut {NoStop}%
\bibitem [{\citenamefont {Wells}(1996)}]{WellsThesis}%
  \BibitemOpen
  \bibfield  {author} {\bibinfo {author} {\bibfnamefont {J.-P.~R.}\
  \bibnamefont {Wells}},\ }\emph {\bibinfo {title} {Laser Spectroscopy of
  Alkaline Earth Fluoride Crystals Doped with Trivalent Samarium and Europium
  Ions}},\ \href@noop {} {Ph.D. thesis},\ \bibinfo  {school} {University of
  Canterbury} (\bibinfo {year} {1996})\BibitemShut {NoStop}%
\bibitem [{\citenamefont {Wells}\ \emph {et~al.}(1999)\citenamefont {Wells},
  \citenamefont {Yamaga}, \citenamefont {Han}, \citenamefont {Gallagher},\ and\
  \citenamefont {Honda}}]{Wells1999}%
  \BibitemOpen
  \bibfield  {author} {\bibinfo {author} {\bibfnamefont {J.-P.~R.}\
  \bibnamefont {Wells}}, \bibinfo {author} {\bibfnamefont {M.}~\bibnamefont
  {Yamaga}}, \bibinfo {author} {\bibfnamefont {T.~P.~J.}\ \bibnamefont {Han}},
  \bibinfo {author} {\bibfnamefont {H.~G.}\ \bibnamefont {Gallagher}}, \ and\
  \bibinfo {author} {\bibfnamefont {M.}~\bibnamefont {Honda}},\ }\href
  {\doibase 10.1103/PhysRevB.60.3849} {\bibfield  {journal} {\bibinfo
  {journal} {Phys. Rev. B}\ }\textbf {\bibinfo {volume} {60}},\ \bibinfo
  {pages} {3849} (\bibinfo {year} {1999})}\BibitemShut {NoStop}%
\bibitem [{\citenamefont {Yamaga}\ \emph {et~al.}(2000)\citenamefont {Yamaga},
  \citenamefont {Honda}, \citenamefont {Wells}, \citenamefont {Han},\ and\
  \citenamefont {Gallagher}}]{Wells2000}%
  \BibitemOpen
  \bibfield  {author} {\bibinfo {author} {\bibfnamefont {M.}~\bibnamefont
  {Yamaga}}, \bibinfo {author} {\bibfnamefont {M.}~\bibnamefont {Honda}},
  \bibinfo {author} {\bibfnamefont {J.-P.~R.}\ \bibnamefont {Wells}}, \bibinfo
  {author} {\bibfnamefont {T.~P.~J.}\ \bibnamefont {Han}}, \ and\ \bibinfo
  {author} {\bibfnamefont {H.~G.}\ \bibnamefont {Gallagher}},\ }\href
  {http://stacks.iop.org/0953-8984/12/i=40/a=314} {\bibfield  {journal}
  {\bibinfo  {journal} {Journal of Physics: Condensed Matter}\ }\textbf
  {\bibinfo {volume} {12}},\ \bibinfo {pages} {8727} (\bibinfo {year}
  {2000})}\BibitemShut {NoStop}%
\bibitem [{\citenamefont {Macfarlane}\ and\ \citenamefont
  {Shelby}(1984)}]{Macfarlane1984}%
  \BibitemOpen
  \bibfield  {author} {\bibinfo {author} {\bibfnamefont {R.~M.}\ \bibnamefont
  {Macfarlane}}\ and\ \bibinfo {author} {\bibfnamefont {R.~M.}\ \bibnamefont
  {Shelby}},\ }\href {\doibase 10.1364/OL.9.000533} {\bibfield  {journal}
  {\bibinfo  {journal} {Opt. Lett.}\ }\textbf {\bibinfo {volume} {9}},\
  \bibinfo {pages} {533} (\bibinfo {year} {1984})}\BibitemShut {NoStop}%
\end{thebibliography}

%

\end{document}